# Grain Boundary Segregation Transitions and Critical Phenomena in Binary Regular Solutions: A Systematics of Complexion Diagrams with Universal Characters

Naixie Zhou, Chongze Hu, Jian Luo[*]

Department of NanoEngineering; Program of Materials Science and Engineering

University of California, San Diego

## Abstract

A systematics of grain boundary (GB) segregation transitions and critical phenomena has been derived to expand the classical GB segregation theory. Using twist GBs as an example, this study uncovers when GB layering *vs.* prewetting transitions should occur and how they are related to one another. Moreover, a novel descriptor, normalized segregation strength ($\phi_{seg}$), is introduced. It can represent several factors that control GB segregation, including strain and bond energies, as well as misorientation for small-angle GBs (in a mean-field approximation), which had to be treated separately in prior models. In a strong segregation system with a large $\phi_{seg}$, first-order layering transitions occur at low temperatures and become continuous above GB roughing temperatures. With reducing $\phi_{seg}$, the layering transitions gradually merge and finally lump into prewetting transitions without quantized layer numbers, akin to Cahn's critical-point wetting model. Furthermore, GB complexion diagrams with universal characters are constructed as the GB counterpart to the classical exemplar of Pelton-Thompson regular-solution binary bulk phase diagrams.

[*] Corresponding author. E-mail: jluo@alum.mit.edu

# Graphical Abstract

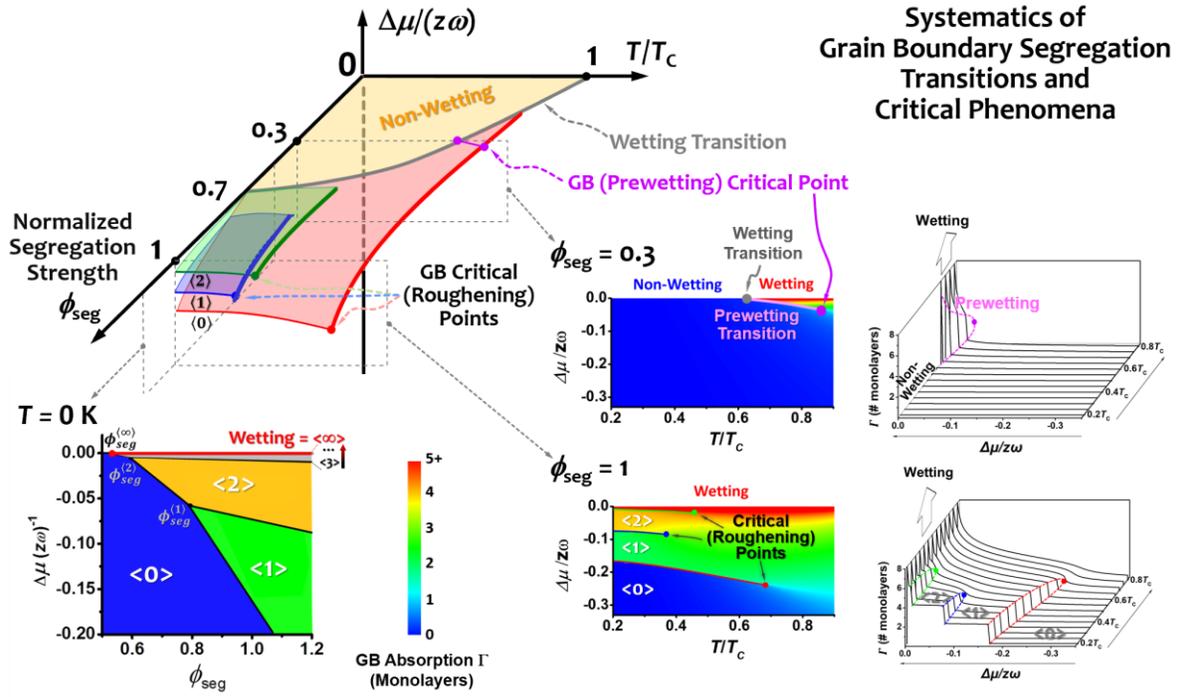

# 1. Introduction

Grain boundary (GB) segregation (*a.k.a.* adsorption in thermodynamics) is an important phenomenon for materials science because it can critically influence microstructural evolution and a broad range of materials properties [1, 2]. Over decades, several classical statistical thermodynamic models have been developed. In 1957, McLean proposed a GB segregation model [3] analogous to the Langmuir surface adsorption model [4], which considered fixed segregation sites and assumed ideal adsorbate-adsorbate interactions (here, GB adsorbates = segregating solute atoms). Various researchers have studied GB segregation experimentally and proposed GB segregation models [5-12]. Notably, Hondros and Seah [13] suggested the existence of first-order segregation transition with strong adsorbate-adsorbate attraction at GBs, in an analogy to the Fowler-Guggenheim surface adsorption model [14]; this segregation transition occurs when the effective GB regular-solution parameter is >$2R$ ($R$ is gas constant), thereby suggesting a GB "phase separation" [15]. More rigorous models of GB segregation transitions have been proposed recently. On the one hand, analogous to Cahn's critical-point wetting model [16], Wynblatt and Chatain predicted a prewetting transition using a lattice adsorption model [17, 18]; it should be noted that similar coupled GB prewetting and premelting transitions had been suggested by phase-field (diffuse-interface) models by Tang *et al.* [19] and Mishin *et al.* [20]. On the other hand, Rickman *et al.* suggested first-order GB layering transitions (for a system resembling Cu-Ag) mostly based on elastic interactions in a micromechanical framework [21]. However, it is not yet known when GB prewetting *vs.* layering transitions should occur and how they relate to each another, which represents the first motivation of this study.

In a broader context, Hart first proposed to treat GBs as 2D interfacial phases in 1968 [22], which were more recently named as "complexions" [15, 23-25] to differentiate them from thin layers of bulk (3D) phases precipitated at GBs. Notably, Dillon *et al.* discovered a series of six discrete GB complexions with increasing total adsorption levels in $Al_2O_3$ based ceramics [15, 26, 27]. These Dillon-Harmer complexions [15, 26-28] resemble the characters of layering and prewetting transitions. However, they were found in ceramic systems with different dopants; it is yet unknown when particular types of complexions should form. Moreover, ceramic GBs are generally complex with interfacial electrostatic and London dispersion interactions [15, 29-31]. Various GB complexions and segregation-induced GB transitions have also been observed in simpler metallic alloys [32-41]. In addition to the lattice [18, 42] and phase-field [19, 20] models, the occurrences of first-order GB transitions have been suggested by atomistic simulations [43-49] and machine learning [50], and experimentally evident in Si-Au [51]. Recently, GB structural transition in elemental Cu [52] and GB topological phase transitions [53] have also been revealed. In a broader contents, GB complexions and transitions have observed or suggested in various binary alloys [32-37, 40, 47, 51, 54-57] and multicomponent (or even high-entropy) alloys [38, 58-64]. However, prior studies of GB transitions are mostly limited to isolated cases and specific systems. This study further reveals a systematics of GB segregation transitions and critical phenomena in binary regular-solution alloys.

GB transitions can cause abrupt changes in microstructural evolution and various materials properties [15, 25, 29]. Here, a potential transformative research direction is represented by developing the GB counterparts to the bulk phase diagrams. To date, such GB "phase" or complexion diagrams have been constructed only for a limited number of specific systems [19, 20, 29, 45, 46, 65-72]. In the history of developing bulk phase diagrams, Pelton and Thompson used two regular solutions to construct a matrix of binary phase diagrams that exhibit most "topological features" commonly seen in real binary alloy phase diagrams, which became a textbook paradigm [73]. This study further aims to establish GB counterparts to the Pelton-Thompson exemplar of regular-solution phase diagrams.

## 2. The Generic Model and Methods

We should first note that lattice models have been used to describe GB segregation (including segregation transitions) by various researchers, including Kikuchi and Cahn [74], Wynblatt and Chatain [17, 18], and Rickman *et al.* [21]. Lattice models have also been used by Trelewicz and Schuh [75] and Darling *et al.* [76], among others [77-79], to model the stability of nanocrystalline alloys (with GB segregation).

In this work, we first show a general framework where different GB potentials can be plugged in to represent different lattice type models. Specifically, we derive a GB potential following the Wynblatt-Chatain model [17] and utilize it for the majority of our analysis (noting that Wynblatt and Chatain's prior analysis of this model [18] revealed a prewetting transition, but not the layering transitions). Other types of GB potentials can also be used in this general framework. This model can also be used to reproduce the layering transitions shown for a Cu like alloy by Rickman *et al.*

[21].

Notably, this model uncovers the new physics of GB segregation transitions by revealing the evolution from the layering to prewetting transition and how it depends on a small number of (mainly three) normalized thermodynamic variables. Moreover, we aim at establishing a unified and general framework to understand GB segregation transitions as a basis to develop GB complexion diagrams.

*2.1 An Ising-Type Model for GB Segregation*

Here, we adopt an Ising-type lattice model to represent the segregation near twist GBs (to represent either general or small-angle GBs) in a binary A-B regular solution. As shown in Fig. 1(a), each lattice site is represented by a "spin" variable $n_k = 0$ (or 1) for the occupation of a solvent atom A (or a solute atom B). The flow chart of the model development is shown in Fig. 1(b). The Hamiltonian of a microstate is expressed as:

$$\mathcal{H} = \sum_k V_k \cdot n_k + \frac{1}{2} \sum_k \sum_{l \neq k} J_{kl}, \tag{1}$$

where $V_k$ represents the potential energy and $J_{kl}$ is the Ising-type pairwise interaction: $J_{kl} = \omega \equiv [e_{AB} - (e_{AA} + e_{AB})/2)]$, where $e_{AA}$, $e_{BB}$, and $e_{AB}$ are the bond energies (taken to be negative values), if $k$ and $l$ are the nearest neighbors and $n_k \neq n_l$; otherwise, $J_{kl} = 0$.

For a binary A-B system, the GB energy ($\gamma_{GB}$) is the GB excess of grand potential. The grand potential is defined as $\Phi_G \equiv U - TS - (\mu_A N_A + \mu_B N_B)$, where $U$ is internal energy, S is entropy, $\mu_A$ and $\mu_A$ are chemical potentials (per atom), and $N_A$ and $N_B$ are total numbers of A and B atoms, respectively. For a symmetric twist GB, the (relative) change in the grand potential per unit area after adding the solute B atoms can be expressed using the Bragg-Williams approximation [80] in a mean-field approach as:

$$\Delta \Phi_G / A_{GB} = 2N_0 \left[ X_i \cdot V_i + \overline{J}_i - \overline{S}_i T - (1 - X_i) \cdot \mu_A - X_i \cdot \mu_B \right], \tag{2}$$

where $X_i$ is the solute fraction on the $i^{th}$ layer from the GB core, $V_i$ represents the GB potential on the $i^{th}$ layer (assumed to be identical within the same layer in the mean-field model), $A_{GB}$ is the GB area, $N_0$ is the number of the lattice points per unit area in each atomic layer, and the prefactor 2 corresponds to the summation of two sides of a symmetric twist GB. Here, $\overline{J}_i$ and $\overline{S}_i$ (defined below) are the averaged pair-interaction energy and mixing configurational entropy, respectively, per atom on the $i^{th}$ layer. We neglect the difference in the vibrational entropy before and after mixing (by assuming them to be identical). Since we need to find the $X_i$ profile that minimizes the grand potential, it is convenient to rewrite Eq. (2) and define a relative grand potential per unit area (by removing the $X_i$-independent constant terms) as:

$$\overline{\Omega} = 2N_0 \sum_i \left[ X_i(V_i - \Delta\mu) + \overline{J}_i - \overline{S}_i T \right] = \Delta\Phi_G / A_{GB} + \text{constant} , \qquad (3)$$

where $\Delta\mu = \mu_B - \mu_A$ is the chemical potential difference (upon replacing a solvent A atom with a solute B atom). For convenience, we can select the reference states so that $\Delta\mu = 0$ at the bulk two-phase co-existence for our numerical calculations. Here, the averaged pairwise interaction energy (*i.e.*, the bond energies per atom referenced to the pure elements) can be expressed as:

$$\overline{J}_i = \omega(z - 2z_v)(1 - X_i)X_i + \tfrac{1}{2}z_v\omega\left[(1 - X_{i-1})X_i + X_{i-1}(1 - X_i) + (1 - X_{i+1})X_i + X_{i+1}(1 - X_i)\right], \qquad (4)$$

where $z$ the coordination number (number of bonds per atom) and $z_v$ is the number of bonds per atom between two adjacent layers (so that $(z - 2z_v)$ is the number of bonds within the plane). In Eq. (4), the first term represents the interaction between atoms within the same layer and the second term corresponds to the interaction between adjacent layers. From Eq. (4), we can further obtain:

$$\overline{J}_i = z\omega(1 - X_i)X_i + \tfrac{1}{2}z_v\omega(X_i - X_{i+1})^2 + \tfrac{1}{2}z_v\omega(X_i - X_{i-1})^2 , \qquad (5)$$

where the first term is the homogenous interaction energy (corresponding the layer composition $X_i$) and the two other terms represent the extra energy due to compositional gradients.

The ideal mixing configurational entropy (per atom) on the $i^{th}$ layer is:

$$\overline{S}_i = -k_B\left[X_i \ln X_i + (1 - X_i)\ln(1 - X_i)\right], \qquad (6)$$

where $k_B$ is the Boltzmann constant. By combining Eqs. (1-6), we can obtain:

$$\overline{\Omega} = 2N_0 \sum_i \left\{ (V_i - \Delta\mu)X_i + \omega\left[z(1 - X_i)X_i + z_v(X_i - X_{i+1})^2\right] + k_B T\left[X_i \ln X_i + (1 - X_i)\ln(1 - X_i)\right]\right\}. \qquad (7)$$

In the above equation, the first term represents the change in the grand potential upon adding solute B atoms, the second term arises from the pair interaction, and the third term is the ideal configurational entropy of mixing. Only $X_i$-dependent terms are kept in this referenced $\overline{\Omega}$.

At a given temperature $T$ and bulk composition $X_{\text{bulk}} = X_\infty$ (that defines $\Delta\mu$), the equilibrium composition profile ($X_i$) that minimizes GB energy ($\gamma_{GB} = \overline{\Omega}$ + constant) can be obtained by taking $\partial\overline{\Omega}/\partial X_i = 0$, leading to a set of McLean-type equations:

$$\begin{cases} \dfrac{X_i}{1-X_i} = \dfrac{X_\infty}{1-X_\infty}\exp\left(-\dfrac{\Delta H_i^{seg}}{k_B T}\right) \\ \Delta H_i^{seg} = V_i + 2z\omega\left[(X_\infty - X_i) + \left(\dfrac{z_v}{z}\right)(X_i - 2X_{i+1} - X_{i-1})\right] \end{cases} \quad (8)$$

Here, $\Delta H_i^{seg}$ is the segregation energy of solute B in the $i^{\text{th}}$ layer (noting $X_0 = X_1$ for the twist GB).

## 2.2 The GB Potential

A merit of this Ising-type model is that it can serve as a generic platform to represent various GB segregation models with the choice of the different GB potential functions. In this section, we derive a GB potential based on the Wynblatt-Chatain model [18]. This will be used as the primary example for a systematic numerical analysis of GB segregation transitions and critical phenomena in this work. Similar GB potentials can also be derived for other models. Several examples are given in Table 1.

Let us consider a lattice model with broken bonds crossing the GB plane. The GB energy for a symmetric twist GB can be written as:

$$\gamma_{GB} = \min\{U^{ex} - TS^{ex} - \mu_A \Gamma_A - \mu_B \Gamma_B\} = \min\{U^{ex} - TS^{ex} - \Delta\mu \cdot \Gamma_B\}, \quad (9)$$

where $U^{ex}$, $S^{ex}$, and $\Gamma_B$ ($= -\Gamma_A$ in this lattice model) are the GB excesses of internal energy, entropy, and solute B, respectively. Assuming (for simplicity) that bonds can only form between the adjacent layers across the twist plane (corresponding to the cases with $J_{\max} = 1$ in Refs. [18, 77]), we can derive:

$$U^{ex} = (1-P)z_v N_0\left[-X_1 e_{BB} - (1-X_1)e_{AA} - 2X_1(1-X_1)\omega\right] - 2N_0 \sum_i X_i \Delta E_{el}^i$$
$$+ 2N_0\omega\sum_i\left[zX_i(1-X_i) + z_v(X_i - X_{i+1})^2 - zX_\infty(1-X_\infty)\right] \quad (10)$$

In the above equation, $(1 - P)$ is the fraction of broken bonds across the twist plane on each side. In such a lattice model, the (0 K) interfacial energies for a surface or a GB of pure A are:

$$\gamma_{\text{Surface}}^{(0)} = \tfrac{1}{2} z_v N_0 (-e_{AA}) \tag{11}$$

and

$$\gamma_{\text{GB}}^{(0)} = (1-P) z_v N_0 (-e_{AA}) \tag{12}$$

In Eq. (10), $\Delta E_{el}^i$ is the strain energy for a solute atom on the $i^{th}$ layer. According to the Friedel model [81], the strain energy decreases with the distance to the GB core following an relation derived for twist GBs with two identical terminal orientations [82]:

$$\Delta E_{el}^i = \beta(1-P)\Delta E_{el}^{\infty} \exp\left[-1.01\left(\frac{h^i}{r_B}\right)^{1.53}\right] = \Delta E_{el}^1 \exp\left[-1.01\left(\frac{h^i}{r_B}\right)^{1.53}\right], \tag{13}$$

where $h^i$ is the distance of the $i^{th}$ layer from the plane $i = 1$, $r_B$ (or $r_A$) is the atomic radius of B (or A), $\beta$ is an adjustable parameter, and

$$\Delta E_{el}^{\infty} = \frac{24\pi G_A K_B r_A r_B (r_A - r_B)^2}{4 G_A r_A + 3 K_B r_B}. \tag{14}$$

Here, $K_B$ is the bulk modulus of B and $G_A$ is the shear modulus of A. See Ref. [82] for elaboration.

We also have the expressions for the (ideal) GB excess entropy of mixing:

$$S^{ex} = k_B T \left[X_i \ln X_i + (1-X_i)\ln(1-X_i)\right] - k_B T \left[X_\infty \ln X_\infty + (1-X_\infty)\ln(1-X_\infty)\right] \tag{15}$$

and the GB adsorption (GB excess of solute):

$$\Gamma (= \Gamma_B = -\Gamma_A) = N_0 \sum_i (X_i - X_\infty) \tag{16}$$

We can plug in Eqs. (10), (15) and (16) into Eq. (9) and rearrange/simplify it to obtain:

$$\gamma_{\text{GB}} = \operatorname*{argmin}_{X_i (i=1,2,\ldots)} \left\{ \begin{array}{l} -(1-P)z_v N_0 \left[2(1-X_1)\omega + (e_{BB}-e_{AA})\right] X_1 - 2N_0 \sum_i \Delta E_{el}^i X_i - \Delta\mu \cdot 2N_0 \sum_i X_i \\ +2N_0 \sum_i \omega \left[z X_i (1-X_i) + z_v (X_i - X_{i+1})^2\right] \\ +2N_0 \sum_i k_B T \left[X_i \ln X_i + (1-X_i)\ln(1-X_i)\right] \end{array} \right\} + \text{constant} \tag{17}$$

Comparing Eq. (17) with Eq. (7), we find the following relations:

$$\begin{cases} V_1 = -\left[\tfrac{1}{2}(1-P)z_v(e_{BB}-e_{AA}) + \Delta E_{el}^1 + (1-P)z_v(1-X_1)\omega\right] \\ V_{i>1} = -\Delta E_{el}^i \end{cases} \quad (18)$$

Here, Eq. (18) defines the GB potential for the Wynblatt-Chatain model [18].

Other forms of GB potential functions (see, *e.g.*, Table 1) can also be adopted, which will produce similar general trends. More complicated GB potential functions can be developed in future studies to represent different types of GBs (*e.g.*, asymmetric GBs) and the effects of GB structural changes (with the introduction of additional structural order parameters). In this study, we adopt Eq. (18) for symmetric twist GBs to establish a baseline of GB segregation transitions and critical phenomena in this regular-solution type GB segregation model.

### *2.3. Normalization of the Model*

We define a normalized segregation strength as:

$$\phi_{seg} \equiv -2V_1/(z\omega) . \quad (19)$$

We further define a dimensionless parameter to characterize the decaying of the GB potential:

$$v_i \equiv V_i/V_1 . \quad (20)$$

By the definition: $v_1 = 1$ and $v_{i\to\infty} \to 0$. We note that Eqs. (19) and (20) are general definitions that are applicable to any GB potentials. In this study, we focus on the region of $\omega > 0$, where the critical temperature for the bulk phase separation is:

$$T_C = z\omega/(2k_B), \quad (21)$$

Then, we can rewrite Eq. (7) by using dimensionless parameters $\phi_{seg}$, $v_i$, and $z_v/z$, and normalized thermodynamic variables $T/T_C$ and $\Delta\mu/(z\omega)$, as:

$$\left(\frac{\overline{\Omega}}{z\omega N_0}\right) = \sum_i \left\{ \left[-\phi_{seg}v_i - 2\left(\frac{\Delta\mu}{z\omega}\right)\right]X_i + 2(1-X_i)X_i + 2\left(\frac{z_v}{z}\right)(X_i - X_{i+1})^2 + \left(\frac{T}{T_C}\right)[X_i \ln X_i + (1-X_i)\ln(1-X_i)]\right\} \quad (22)$$

In Eq. (18), the GB potential of the first layer ($V_1$) depends on the composition of that layer ($X_1$) so it is not a constant. However, the third term in Eq. (18) is usually significantly smaller than the differential bonding and strain energies. Thus, we can neglect this $X_1$-dependent term (for simplicity) to define a normalized segregation strength $\phi_{seg}$ based on Eq. (18), as:

$$\phi_{seg} \approx \left[(1-P)z_v(e_{BB}-e_{AA}) + 2\Delta E_{el}^1\right]/(z\omega). \quad (23)$$

For an average general GB in a pure metal, the following empirical relation is well known:

$$\gamma_{GB}^{(0)} \approx \tfrac{1}{3} \gamma_{Surface}^{(0)} \tag{24}$$

Combining Eqs. (11), (12), (23) and (24), we can obtain and adopt the following expression to represent an "average" general twist GB:

$$\begin{cases} 1 - P = \tfrac{1}{6} \\ \Delta E_{el}^1 = \tfrac{1}{6} \beta \Delta E_{el}^\infty \\ \phi_{seg} \approx \left[ \tfrac{1}{6} z_v (e_{BB} - e_{AA}) + 2\Delta E_{el}^1 \right] / (z\omega) \end{cases} \tag{25}$$

An expression of $\phi_{seg}$ for small-angle GBs as a function of misorientation angle ($\Delta\theta$) is derived later and given in Eq. (36). Subsequently, we will show that this normalized segregation strength $\phi_{seg}$ dominates the GB adsorption behaviors.

We further define the ratio of the strain energy contribution to $\phi_{seg}$ as:

$$f_{strain} \equiv \frac{2\Delta E_{el}^1 / (z\omega)}{\phi_{seg}} = \frac{2\beta \Delta E_{el}^\infty}{z_v (e_{BB} - e_{AA}) + 2\beta \Delta E_{el}^\infty}. \tag{26}$$

Combining Eqs. (13), (20), and (26), we can obtain and adopt the following expression for a (100) twist GB of a hypothetic lattice constant 0.361 nm (as an example for our subsequent analytical and numerical analyses):

$$v_i \approx f_{strain} \exp\left[-1.21 \cdot (i-1)^{1.53}\right]. \tag{27}$$

Fig. S1 in the Supplementary Material (SM) compares three different GB potentials based on the Wynblatt-Chatain model [18] using different $f_{strain}$ values, along with the Rickman *et al.*'s model [21, 83]. We will show that $f_{strain}$ is the second most important material parameter (after $\phi_{seg}$) that affect the GB adsorption and transition behaviors.

We again note that the current model is generic, where different forms of the GB potentials can be adopted. This regular-solution type GB model and the complexion diagrams derived in the subsequent section provide a basis to understand the systematics of the GB segregation transitions and critical phenomena, akin to their bulk counterparts of regular solutions and the Pelton-Thompson phase diagrams [73].

## 3. Analytical and Numerical Results

### 3.1 Grand States of GB Adsorption

Let us first analyze the ground states at $T = 0$ K. We use a non-negative integer $n$ to represent the number of adsorption layers on one side of the GB (so that the absorption $\Gamma = 2nN_0$ for the symmetric twist GB for a perfect "quantum" state at 0K). For $n = 0$, $\overline{\Omega}^{\langle 0 \rangle} = 0$ represents a GB state without segregation. At $T = 0$ K, we can simplify Eq. (22) for the $n^{th}$ GB ground state ($n \geq 1$), denoted as "$\langle n \rangle$", as:

$$\frac{\overline{\Omega}^{\langle n \rangle}}{z \omega N_0} = 2\left(\frac{z_v}{z}\right) - \phi_{seg} \sum_{i=1}^{n} v_i - 2n\left(\frac{\Delta \mu}{z \omega}\right) \tag{28}$$

Fig. 2(a) shows the ground states of GB adsorption derived for a representative general GB at $T = 0$ K ($z_v/z = 1/3$; using $v_i$ expressed in Eq. (27) with $f_{strain} = 0.5$ as an example).

As we will show subsequently, an analytical solution can be solved rigorously at 0 K (without entropic effects), which can also set a baseline for, and reflect the behaviors of, strong, medium, and weak segregation systems at high temperatures. The latter (the finite-temperature cases) can only be solved numerically.

The GB adsorption behaviors at 0 K can be classified into three regions based on the value of normalized segregation strength ($\phi_{seg}$). For a strong segregation system ($\phi_{seg} > \phi_{seg}^{\langle 1 \rangle}$, where $\phi_{seg}^{\langle 1 \rangle}$ is defined in Eq. (30) below and shown in Fig. 2(a)), the GB undergoes a complete sequence of adsorption states of $n = 0, 1, 2, ...$ with increasing $\Delta \mu$, until complete wetting ($n = \infty$) occurs at $\Delta \mu = 0$. Fig. 2(b) schematically illustrates $\overline{\Omega}^{\langle n \rangle}$ vs. $\Delta \mu/(z\omega)$ lines. The transition between the $\langle n \rangle$ and $\langle n + 1 \rangle$ states take place when $\overline{\Omega}^{\langle n \rangle} = \overline{\Omega}^{\langle n+1 \rangle}$, at the normalized chemical potential:

$$\begin{cases} \left(\dfrac{\Delta \mu}{z \omega}\right)_{\langle 0 \leftrightarrow 1 \rangle} = \left(\dfrac{z_v}{z}\right) - \tfrac{1}{2} \phi_{seg} \\ \left(\dfrac{\Delta \mu}{z \omega}\right)_{\langle n \leftrightarrow n+1 \rangle} = -\tfrac{1}{2} \phi_{seg} v_{i+1} \quad (n \geq 1) \end{cases} \tag{29}$$

Here, a larger $\phi_{seg}$ enables the transition to occur at a more negative $\Delta \mu/(z\omega)$ (lower $\mu_B$).

For a medium segregation system ($\phi_{seg}^{\langle 1 \rangle} \geq \phi_{seg} > \phi_{seg}^{\langle \infty \rangle}$, as shown in Fig. 2(a)), an incomplete sequence of layering transitions occurs, where the $\langle 0 \rangle$ state transits to the $\langle n + 1 \rangle$ state directly (in the absence of the $\langle 1 \rangle$ to $\langle n \rangle$ states). This threshold (triple-phase point) can be solved by letting $\overline{\Omega}^{\langle 0 \rangle} = \overline{\Omega}^{\langle n \rangle} = \overline{\Omega}^{\langle n+1 \rangle} = 0$, as:

$$\phi_{seg}^{\langle n \rangle} = 2\left(\frac{z_v}{z}\right) \cdot \left(\sum_{i=1}^{n} v_i - n \cdot v_{n+1}\right)^{-1} \tag{30}$$

The $\langle 0 \leftrightarrow n+1 \rangle$ "skip" transition line ($n \geq 1$) is solved by letting $\overline{\Omega}^{\langle 0 \rangle} = \overline{\Omega}^{\langle n+1 \rangle} = 0$, as:

$$\left(\frac{\Delta \mu}{z\omega}\right)_{\langle 0 \leftrightarrow n+1 \rangle} = \frac{1}{n+1} \cdot \left[\left(\frac{z_v}{z}\right) - \tfrac{1}{2} \phi_{seg} \sum_{i=1}^{n+1} v_i \right] \tag{31}$$

For weak segregation systems ($\phi_{seg} < \phi_{seg}^{\langle \infty \rangle}$), the $\langle 0 \rangle$ state is stable at all chemical potentials at 0 K. This threshold can be obtained by letting $n \to \infty$ in Eq. (30), as:

$$\phi_{seg}^{\langle \infty \rangle} = 2\left(\frac{z_v}{z}\right) \cdot \left(\sum_{i=1}^{\infty} v_i\right)^{-1}. \tag{32}$$

Eqs. (29)-(32) can be used to construct GB segregation diagrams at 0 K. One primary example is shown in Fig. 2(a) for an average general (100) twist GB in an FCC metal ($z_v/z = 1/3$) with $f_{strain} = 0.5$ from the Wynblatt-Chatain model [18]. More instances derived from different GB potential functions can be found in Suppl. Fig. S2 in the SM for different models or systems.

### 3.2 Systematics of Complexion Diagrams for Average General GBs

With a clear understanding of the ground states at 0 K, we can now investigate the GB segregation transitions and critical phenomena at finite temperatures. We can express our model with five independent dimensionless variables or parameters:

(a) normalized temperature $T/T_C$,

(b) normalized chemical potential difference $\Delta \mu/(z\omega)$ (dictated by $X_{bulk} = X_\infty$),

(c) normalized segregation strength $\phi_{seg}$,

(d) fraction of the strain contribution $f_{strain}$, which determines the decaying of GB potential $v_i$ in the Wynblatt-Chatain model [18] according to (Eq. 27), and

(e) the fraction of the one-side out-of-plane bonds $z_v/z$ (which is a crystallographic parameter dependent on the crystal structure and the orientation of the twist plane).

Here, (a) and (b) are thermodynamic variables; (c) is the dominating, while (d) and (e) are secondary, material or crystallographic parameters. In the numerical analysis, we adopt Eqs. (25) and (27) as the GB potential for an "average general GB" ($1 - P = 1/6$ and $z_v/z = 1/3$ to represent general (100) twist GBs in FCC alloys, unless noted otherwise).

By solving the equilibrium $X_i$ profile via minimizing $\bar{\Omega}$ in Eq. (7) (or the normalized Eq. (22)), we can compute GB adsorption $\Gamma$ (Eq. (16)) as a function of two thermodynamic variables ($T/T_C$ and $X_{\text{bulk}}$ or $\Delta\mu/(z\omega)$) for given material parameters ($\phi_{seg}$, $f_{\text{strain}}$, and $z_v/z$).

Fig. 3 compares the typical computed $\Gamma$ vs. $X_{\text{bulk}}$ ($= X_\infty$) at different $T/T_C$'s of strong vs. weak segregation systems (for a representative case of $f_{\text{strain}} = 0.5$). On the one hand, a strong segregation system ($\phi_{seg} = 1$) shows a complete series of first-order layering transitions with discontinuous increases in $\Gamma$ (Fig. 3(b)), mimic its 0 K character. Here, the $\Gamma$ in the $\langle n \rangle$ state corresponds to nominally $2n$ monolayers (albeit not exact integer numbers at $T > 0$ K due to entropic effects) for the twist GB. With increasing temperature, each first-order transition vanishes at a GB critical (roughening) temperature ($T_{\text{GB crit.}}^{\langle n \leftrightarrow n+1 \rangle}$), above which the transition becomes continuous (Fig. 3(b)). With increasing $X_{\text{bulk}}$, $\Gamma$ is divergent ($n \to +\infty$) at $\Delta\mu = 0$ to (complete) wetting at any temperature.

On the other hand, a weak segregation system ($\phi_{seg} = 0.4$) shows a single prewetting (thin-thick) segregation transition vanishing at a <u>p</u>re<u>w</u>etting <u>c</u>ritical (PWC) point at $T_{\text{GB crit.}}^{PWC}$, where the equilibrium segregation states do not correspond to any integer numbers of monolayers (Fig. 3(d)). This case is analogous to the Cahn critical-wetting model [16, 18]. In the corresponding $T/T_C$-$X_{\text{bulk}}$ GB segregation diagram (Fig. 3(e)), the prewetting transition line starts (at the low-temperature end) from $T_W$ (on the boundary of the bulk miscibility gap, which corresponds to a first-order wetting transition, i.e., wetting at $T > T_W$, in the phase separation region), and it terminates (at the high-temperature end) at $T_{\text{GB crit.}}^{PWC}$.

Next, we represent the GB segregation transitions in a 3D diagram as a function of $T/T_C$, $\phi_{seg}$, and $\Delta\mu/z\omega$ (Fig. 4(a)), which include GB (layering or prewetting) transition hyper surfaces ending at connected lines of GB critical temperatures, along with a line of the wetting transition on the horizonal plane of $\Delta\mu = 0$. Five selected 2D cross-sectional maps of the $T/T_C$-$\Delta\mu/z\omega$ GB segregation diagrams at $\phi_{seg} = 1, 0.8, 0.7, 0.6$, and $0.3$, respectively, along with the corresponding computed $\Gamma$ vs. $\Delta\mu/z\omega$ curves at different $T/T_C$'s, are plotted in Figs. 4(b-f) and discussed below.

Fig. 4(b) shows the strong segregation system of $\phi_{seg} = 1$ as functions of $T/T_C$ and $\Delta\mu/(z\omega)$ with a complete sequence of layering transitions ($\langle 0 \rangle, \langle 1 \rangle, \langle 2 \rangle, ...$) that ends at a series of critical (roughening) points at $T_{\text{GB crit.}}^{\langle n \leftrightarrow n+1 \rangle}$. This is an alternative (clearer) representation of Fig. 3(b, c). With decreasing $\phi_{seg}$, the $\langle 0 \leftrightarrow 1 \rangle$ layering transition line shifts towards less negative $\Delta\mu$ and merges with the $\langle 1 \leftrightarrow 2 \rangle$ transition line to form a direct $\langle 0 \leftrightarrow 2 \rangle$ transition, as shown in Fig. 4(c) *vs.* 4(d). After the merging, the $\langle 1 \rangle$ state is no longer stable; with increasing temperature, the actual adsorption deviates from the nominal values (*i.e.*, the "quantum" numbers become fuzzy) until the first-order $\langle 0 \leftrightarrow 2 \rangle$ transition ends at a merged GB critical point at $T_{GB\ crit.}^{\langle 0 \leftrightarrow 2 \rangle}$. The complete merging of these two layering transitions at $\phi_{seg}^{\langle 0 \leftrightarrow 2 \rangle}$ corresponds to the onset of the medium segregation region in the GB ground state (at 0 K) shown in Fig. 2(a). With further reduction of $\phi_{seg}$, the GB layering transitions further merge, producing a series of merged transition lines starting from 0K at $\phi_{seg}^{\langle n \rangle}$ and ending at $T_{GB\ crit.}^{\langle 0 \leftrightarrow n+1 \rangle}$. In the weak segregation region ($\phi_{seg} < \phi_{seg}^{\langle \infty \rangle}$), all layering transitions lump into a prewetting transition [16, 18] (Fig. 4(f)). In the 3D adsorption diagram (Fig. 4(a)), the prewetting vanishes at $\phi_{seg}$-dependent $T_{GB\ crit.}^{PWC} = T_{GB\ crit.}^{\langle 0 \leftrightarrow \infty \rangle}$. The prewetting is divergent at $\Delta\mu = 0$ to (complete) wetting.

Fig. 2(a) is the $T = 0$ K cross section of the 3D adsorption diagram shown in Fig. 4(a), which sets the behaviors of strong, medium, and weak segregation systems at the ground states. At higher temperatures, all GB transitions shift towards more negative $\Delta\mu$ due to a competition between entropic and enthalpic effects.

Based on the Landau theory [84], a first-order transition becomes continuous when $d^2\overline{\Omega}_{min}/d\Gamma^2 = 0$, where we can adopt $\Gamma$ as the order parameter for the current case. While a rigorous analytic solution does not exist, we can use an approximation $\Gamma \approx 2[(n-1) + X_n]N_0$ to estimate the GB critical temperatures. Specifically, the approximate critical temperature is estimated by assuming all $X_i = 1$ for $i < n$ layers (but allowing $X_n$ to vary) and letting the second derivative Eq. (22) with respect to $\Gamma$ (or $X_n$ since $\Gamma \approx 2[(n-1) + X_n]N_0$) to zero at $X_n = 0.5$ as:

$$T_{GB\ crit.}^{\langle 0 \leftrightarrow 1 \rangle} \approx \left[1 - (z_v/z)\right]T_C \tag{32}$$

and

$$T_{GB\ crit.}^{\langle n \leftrightarrow n+1 \rangle} \approx \left[1 - 2(z_v/z)\right]T_C \text{ (for } n \geq 1\text{)}. \tag{33}$$

In the current example, $z_v/z = 1/3$ so that $T_{crit}^{\langle 0 \leftrightarrow 1 \rangle} \approx 0.67T_C$ and $T_{crit}^{\langle 1 \leftrightarrow 2 \rangle} \approx 0.33T_C$, which agree with the actual values shown in Fig. 3 approximately. Noting that the $\langle 0 \leftrightarrow 1 \rangle$ transition and $T_{GB\ crit.}^{\langle 0 \leftrightarrow 1 \rangle}$ differ from the subsequent $\langle n \leftrightarrow n+1 \rangle$ transition and $T_{\text{GB crit.}}^{\langle n \leftrightarrow n+1 \rangle}$ for $n \geq 1$ since $\langle 0 \leftrightarrow 1 \rangle$ is the first transition that occurs without a preexisting adsorption layer. Specifically, the $\langle 1 \leftrightarrow 2 \rangle$ roughening may be easier to occur than the $\langle 0 \leftrightarrow 1 \rangle$ roughening because steps and kinks can develop on the preexisting $\langle 1 \rangle$ adsorption layer. This explains that $T_{\text{GB crit.}}^{\langle 0 \leftrightarrow 1 \rangle} > T_{\text{GB crit.}}^{\langle 1 \leftrightarrow 2 \rangle}$. More accurate numerical results show $T_{\text{GB crit.}}^{\langle 1 \leftrightarrow 2 \rangle} < T_{\text{GB crit.}}^{\langle 2 \leftrightarrow 3 \rangle}$, at least for the two cases $f_{\text{strain}} = 0.5$ and 1 (Fig. 3(b), Fig. 4 (b, c), and Suppl. Figs. S5 and S6, as well as "General GBs Examples II & III", in the Supplementary Material). Here, we suspect that the $\langle 2 \leftrightarrow 3 \rangle$ transition is stabilized with a high critical temperature than that of the $\langle 1 \leftrightarrow 2 \rangle$ transition due to a strain effect.

Figs. 3 and 4 were computed for a mixed case of $f_{\text{strain}} = 0.5$ (*i.e.*, the strain and differential bond energies contribute to 50%-50% to $\phi_{seg}$). Fig. 5 (a) *vs.* (b) further shows computed 3D GB segregation transition diagrams for $f_{\text{strain}} = 0$ and $f_{\text{strain}} = 1$, respectively, where segregation is driven solely by the differential bonding or strain energy, respectively. Furthermore, Fig. 6 shows a matrix of nine (3×3) representative 2D GB segregation diagrams for $\phi_{seg} = 1.0, 0.7$, and 0.4, respectively, and $f_{\text{strain}} = 0, 0.5$, and 1, respectively. Additional detailed calculated results are documented in Suppl. Figs. S3-S6 in the SM. All three cases ($f_{\text{strain}} = 0, 0.5$, and 1) show analogous key characteristics of GB adsorption (layering and prewetting) transitions and critical phenomena, while the GBs transitions take place at different $\Delta\mu/z\omega$ values (dependent on $f_{\text{strain}}$). Specifically, the strain energy result in a longer-range GB potential (with slower decaying in $v_i$), so that layering transitions shift towards to more negative $\Delta\mu/z\omega$ with a larger $f_{\text{strain}}$.

In addition, the $z_v/z$ ratio can also have some influences. Suppl. Fig. S7 in the SM compares computed complexion diagrams for two general GBs: an FCC (100) twist GB with $z_v/z = 1/3$ *vs.* a simple cubic (100) twist GB with $z_v/z = 1/6$. Both cases share the similar layering transitions and critical phenomena, while the specific positions of the transition lines and critical points are somewhat different. A smaller $z_v/z$ results in a stronger tendency for the occurrence layering transitions, which transit to the prewetting region at a larger $\phi_{seg}$. This can also be evident in the analytical solutions for the 0 K ground states, because the thresholds for the weak and medium segregation regions $\phi_{seg}^{\langle 1 \rangle}$ and $\phi_{seg}^{\langle \infty \rangle}$ both are proportional to $z_v/z$, as shown in Eqs. (30) and (32).

In summary, all the cases discussed in this section exhibit universal characters in computed GB complexion diagrams. These results collectively suggest that $\phi_{seg}$ is the dominant factor controlling the segregation transitions and critical phenomena, while $f_{strain}$ and $z_v/z$ play secondary roles to influence the specific positions of the first-order transition lines and critical points.

*3.3 The Equivalence of General and Small-Angle GBs and Influences of the Misorientation*

In addition to general GBs, this model can also represent small-angle twist GBs following an approach by Wynblatt *et al.* [82]. This is a mean-field approximation to obtain the effective GB potential averaged over the plane parallel to the GB. Specifically, the general and small-angle GBs have virtually identical GB segregation diagrams if the normalized segregation strength $\phi_{seg}$ is identical (achieved by selecting different strain and/or differential bonding energies, as well misorientation angle $\Delta\theta$ for small-angle GBs, to match the same $\phi_{seg}$).

Here, we use small-angle (100) twist GBs in an FCC alloy as an illustrating example. In this mean-field approximation, the relation between twist angle $\Delta\theta$ and broken bond fraction can be expressed as [82]:

$$(1-P) \approx 0.18\sin^{1/4}(2\Delta\theta). \tag{34}$$

Moreover, the elastic energy is also proportional to the GB free volume or $(1 − P)$ [82]:

$$\Delta E_{el}^1 = (1-P)\beta\Delta E_{el}^\infty. \tag{35}$$

Plugging in Eqs. (34) and (35) into Eq. (23), the normalized segregation strength can be expressed as a function of $\Delta\theta$ for a small-angle twist GB as:

$$\phi_{seg} \approx 0.18\sin^{1/4}(2\Delta\theta)\left[z_v(e_{BB}-e_{AA})+2\beta\Delta E_{el}^\infty\right](z\omega)^{-1}. \tag{36}$$

Thus, $\phi_{seg}$ for small-angle GBs can be varied by changing $\Delta\theta$, $z_v(e_{BB}-e_{AA})$, or $\Delta E_{el}^\infty$, with essentially equivalent effects.

Fig. 7 shows four (2×2) computed adsorption diagrams for small-angle GBs (with $f_{strain} = 0$) for $z_v(e_{BB}-e_{AA}) = 3z\omega$ and $6z\omega$, respectively, and $\Delta\theta = 5°$ and $10°$, respectively. In this series, the corresponding $\phi_{seg}$ are computed to be ~0.34, ~0.40, ~0.68, and ~0.81, respectively. With the increasing $\phi_{seg}$, prewetting (in weak segregation systems) to layering (in strong segregation systems) transitions have been observed, akin to those observed in general GBs.

Furthermore, Fig. 8 shows a computed 3D GB segregation transition diagram for a case of $z_v(e_{BB} - e_{AA}) = 6z\omega$ ($f_{strain} = 0$), plotted in the space of $\Delta\theta$, $T/T_C$, and $\Delta\mu/(z\omega)$. More detailed computation results for this case, as well as another case with smaller $z_v(e_{BB} - e_{AA}) = 3z\omega$, can be viewed in a slide show in the (D) and (E) parts (Small-Angle GBs Examples I & II) of the Appendix of the Supplementary Material. If we replot Fig. 8 in the space of $\phi_{seg}$ (with the mathematical transformation of $\phi_{seg} \approx 1.08 \sin^{1/4}(2\Delta\theta)$ following Eq. (38)), $T/T_C$, and $\Delta\mu/z\omega$, it should overlap with (be identical to a part of) the 3D GB segregation transitions diagram shown in Fig. 5(a) with $f_{strain} = 0$.

Fig. 9 shows computed 2D GB complexion diagrams for two series of small-angle GBs (each with different $\Delta\theta$ values but fixed $z_v(e_{BB} - e_{AA})$ or $\beta\Delta E_{el}^{\infty}$) and one series of general GBs (varying $z_v(e_{BB} - e_{AA})$ or $\beta\Delta E_{el}^{\infty}$) to illustrate the equivalence of small-angle and general GBs with identical $\phi_{seg}$. In each row, the misorientation angles ($\Delta\theta$) and other model parameters are selected so that the final $\phi_{seg}$ values are identical for two small-angle GBs and a third general GBs. Each computed GB segregation diagram can represent three equivalent cases (two small-angle GBs and a general GB). These comparisons of three series of equivalent GBs have been conducted for both $f_{strain} = 0$, as shown in Fig. 9(a), and $f_{strain} = 1$, as shown in Fig. 9(a). Again, $f_{strain}$ has a secondary influence here.

## 4. Discussion

### *4.1 Normalized Segregation Strength*

The normalized segregation strength $\phi_{seg}$ defined in this study is a useful descriptor that can consider the combined effects of bonding and strain energies for both general and small-angle GBs, as well as misorientation for small-angle GBs. These factors had to be treated separately in prior studies [17, 18, 21]. Specifically, $\phi_{seg}$ to be the dominating predicting parameter, where we found similar GB segregation transitions and critical phenomena for systems for the same $\phi_{seg}$. In addition, $f_{strain}$ (or $v_i$ in general) and $z_v/z$ have secondary influences.

As we have discussed above, the expressions of $\phi_{seg}$ can also help us to establish equivalence among general GBs and different small-angle GBs. For each small-angle GB, we can find a general GB with greater $z_v(e_{BB} - e_{AA})$ or/and $\beta\Delta E_{el}^{\infty}$ (while fixing $f_{strain}$ and $z_v/z$) to match the same value of $\phi_{seg}$, which will have a virtually identical GB segregation diagram. Eq. (36) also establishes an equivalence between increasing (or decreasing) $\Delta\theta$ and increasing (or decreasing) $z_v(e_{BB} - e_{AA})$ or/and $\beta\Delta E_{el}^{\infty}$ among different small-angle GBs.

We should note that this equivalence is held under the approximations adopted in this mean-field model and affected by the accuracies of GB potentials. For example, for small-angle GBs, periodical GB dislocations will lead to 2D segregation patterns within the GB plane, which is not captured in the mean-field approach. In addition, the GB potential functions adopted are only first-order approximations (which can be improved via developing more realistic GB potentials).

In summary, the normalized segregation strength $\phi_{seg}$ can be used to forecast the systematic trends in GB segregation transitions and critical phenomena. It can help to establish the equivalence between small-angle and general GBs.

## *4.2 GB Counterparts to Pelton-Thompson Phase Diagrams*

In most materials thermodynamic textbooks, we start to construct binary alloy phase diagrams using two (a solid and a liquid) regular solutions, following Pelton and Thompson's foundational article in 1975 [73]. They can produce a systematics of binary regular-solution phase diagrams with most common features in real binary alloy phases. These Pelton-Thompson regular-solution phase diagrams serve as a basis to understand the real alloy phase diagrams that can often be more complex. Moreover, they can forecast useful trends for real alloys as an approximation.

Based on a regular-solution GB model, this study establishes GB counterparts to the Pelton-Thompson regular-solution bulk phase diagrams [73]. Via using normalized variables ($T/T_C$ and $\Delta\mu/z\omega$) and parameters ($\phi_{seg}$, $f_{strain}/v_i$, and $z_v/z$), a systematics of GB segregation complexion diagrams have been computed. Similar to the bulk regular-solution model and Pelton-Thompson phase diagrams [73], this GB model and systematic complexion diagrams can forecast useful trends for various alloys as a regular-solution approximation. Moreover, they serve as the basis for understanding GB segregation transitions and critical phenomena.

The bulk regular-solution model can be subsequently extended to the full <u>cal</u>culation of <u>pha</u>se <u>d</u>iagram (CALPHAD) models to construct more accurate bulk phase diagrams. Likewise, the current regular-solution type GB model and analysis can also be further improved for developing more accurate GB complexions diagrams (as elaborated in §4.6).

*4.3 Comparison with Experiments*

To our best knowledge, no systematic experimental data exist to validate the temperature and chemical potential dependent GB complexion transitions in binary alloys, but several individual cases have been characterized and can be compared with our model. On the one hand, bilayer complexions (*i.e.*, the ⟨1⟩ state) are pervasive in strong segregation systems like Ni-Bi [33, 34], Cu-Bi [85], and Si-Au [51], with $\phi_{seg} > 1$ (estimated to be ~1.3, ~8.3, ~3.1, respectively, based on a Miedema type model [86]). On the other hand, nanometer-thick amorphous-like interfacial films consistent with prewetting layers were found in a weak segregation system Cu-Zr (estimated $\phi_{seg} \approx 0.01$) [36]. These observations support the current model. We also note that W-Ni [37] and Mo-Ni [32], which exhibit coupled GB premelting and prewetting (with experimental characterization of general GBs formed at high temperatures), also have large estimated $\phi_{seg}$ of 1.31 and 3.41, respectively. However, the high-temperature premelting (interfacial disordering) effects (represented in the experiments [32, 37]) are not considered in the current model.

Notably, the current model revealed a systematics of GB segregation transitions and critical phenomena (as discussed above in §4.3). It serves as a basis to develop more realistic GB complexion diagrams, where the effects of structural transitions (reconstruction [34], broken symmetry [46], or disordering [29, 32, 37]) can be further considered.

*4.4 Other GB Segregation Models*

As we have noted earlier, the current lattice GB model can also represent other GB segregation models by selecting different GB potential functions (*e.g.*, those shown in Table 1).

Notably, Rickman *et al.* [21] studied the layering transitions in GB segregation for Read-Shockley type low-angle GBs composed of arrays of dislocations, where GB potential is translational invariant. This current GB model can also represent Rickman *et al.*'s model [21] via adopting a different GB potential function ($v_i = i^{-4}$). This will produce similar general trends of GB segregation transitions and critical phenomena (albeit some differences on the exact positions of transitions and critical points due to the different approximations adopted in the GB potential) and GB complexion diagrams with the same universal characters. The current model can represent not only the general high-angle GB, but also small-angle GBs. In general, we expect similar trends and universal characters in GB segregation diagrams regardless the GB potential adopted, while the specific details can vary (being more or less accurate for different GB segregation models).

*4.5 Comparison with Multilayer Surface Adsorption*

It should be noted that an analogous systematics of layering to prewetting adsorption transitions has been previously derived for multilayer adsorption of inert gas molecules on an attractive substrate by Nakanishi and Fisher [87] and Pandit *et al.* [88]. This study shows parallel adsorption transitions and critical phenomena can exist at internal interfaces for GB segregation in binary alloys, thereby showing generality of interfacial phenomena in apparently different systems.

Suppl. Fig. S8 in the SM compares several models. A classical Ising model for surface absorption (of inert gas) considers only one component, where each lattice site is described as occupied or empty. In the GB (or surface) segregation model for a binary alloy, there are two elements; each site can be occupied by either a solvent or solute atom. In addition, there is only one type of interaction between molecules in physical absorption. For GB segregation, there are three types of stronger, metallic, (A-A, B-B, and A-B) bonds in binary alloys. Interfacial potentials can be used to describe both the surface physical adsorption and GB segregation, but their physical origins, strengths, characteristic length, and scaling laws are all different. The adsorbate-substrate interactions are typically van der Waals or electrostatic forces for surface physical absorption, with possible chemical bonding for the first layer. The GB potential represents the overall effects of bonding and elastic energies, including crystallographic influences, which can be more complex.

Interestingly, analogous interfacial adsorption/segregation transitions and critical phenomena can exist in these apparently different physical systems. While analogous phenomena are well established for surface physical absorption [87, 88], the systematics of the layering and prewetting transitions, roughening, and critical phenomena, along with the universal characters in GB complexion diagrams, are derived for the GB segregation in binary alloys for the first time in this work.

*4.6 Limitations and Generalization of the Current Model*

This current lattice model can be further generalized to represent symmetric tilt or asymmetric

(tilt or mixed) general GBs by introducing different GB potentials, with some additional assumptions. For asymmetric GBs, two separate GB potentials should be introduced to characterize the segregation profiles in the two different lattices, with appropriate boundary conditions to couple the two sides of segregation. In such cases, the layering and prewetting transitions at each side can occur at different chemical potentials, but they should couple and interact one another. Odd numbers of segregation, where are observed in experiments [15, 26, 27], can be produced for either asymmetric GBs or symmetric GBs with an atomic plane at the center. However, simple lattice models do not consider the possible reconstruction of GBs, which can change the symmetry of GBs (*e.g.*, breaking a mirror symmetry [46] or change to different local crystalline structures or 2D symmetries [34, 64]).

The current analysis is based on regular solutions with positive pair-interaction parameters ($\omega >$ 0) where GB segregation transitions will occur. For regular solutions with negative pair-interaction parameters ($\omega < 0$), superlattice ordering and disordering may occur, which represent separation physical phenomena that should be analyzed in a different study.

The current model does not consider the (premelting or liquid-like) interfacial disordering that can often occur at high temperatures [35, 37] and other structural transition (*e.g.*, reconstruction or change of symmetry) [33, 34, 46]. In this regard, the current model treats the chemical (adsorption or segregation) transitions only. Yet, this current study reveals rich new physics of layering and prewetting transitions and related critical phenomena. As we have briefly discussed previously, more complicated GB potential functions may be developed to represent the effects of GB disordering or reconstruction in future studies. In such case, additional order parameters need to be introduced to describe premelting-like interfacial disordering [35, 37] or interfacial reconstruction [33, 34, 46], and the GB segregation potential will depend on such order parameters. These further refinements are feasible, but nontrivial. They will enable further investigation of coupled GB segregation and structural transitions.

## 5. Conclusions

We have derived a systematics of GB segregation transitions and critical phenomena in binary regular-solution alloys. Notably, a normalized segregation strength $\phi_{seg}$ is introduced to represent the overall effects of strain and bond energies for both general and small-angle GBs, as well as the misorientation for small-angle GBs. We showed that strong segregation systems with large $\phi_{seg}$ undergo a series of layering transitions, which gradually merge and lump into a prewetting transition (without quantum numbers) with reducing $\phi_{seg}$. We revealed universal characters in the GB segregation complexion diagrams.

This study has not only revealed a rich spectrum of GB segregation transitions and critical phenomena to enrich the classical GB segregation theory, but also established a generic and extendable model for understanding GB segregation transitions and critical phenomena. Notably, we have established the GB counterparts to the Pelton-Thompson regular-solution bulk phase diagrams [73], which can also serve as a starting point of fundamental importance to systematically understand GB transitions and develop GB complexion diagrams.

**Declaration of Competing Interest**

The authors declare that they have no known competing financial interests or personal relationships that could have appeared to influence the work reported in this paper.

**Acknowledgement**

We acknowledge partial supports by the UCI MRSEC Center for Complex and Active Materials (NSF DMR-2011967, for 2020-2026), a Vannevar Bush Faculty Fellowship (ONR N00014-16-2569, for 2014-2020), and a prior MURI program (ONR N00014-11-1-0678, 2011-2016) for this theoretical work that was initiated in 2013 and has been improved continuously and finally completed over last eight years.

**Supplementary Material (SM)**
The 51-page SM includes:
- Supplementary Figures S1-S8
- <u>Appendix</u>: A Slide Show of Five Groups of 35 Computed GB Diagrams
    - ✓ General GBs Examples I-III (9×3 GB Complexion Diagrams and Related Results)
    - ✓ Small-Angle GBs Examples I and II (8 GB Complexion Diagrams and Related Results)

**Table 1.** Comparison of different GB segregation models. Here, $z_v/z$ is the ratio of bonds connecting one adjacent layer divided by the total coordination number, $\omega$ is the pair-interaction (regular-solution) parameter, $i_{max}$ is the maximum possible number of GB segregation layers considered by the model, and $v_i$ is the normalized GB potential decaying function.

|  | $z_v/z$ | $\omega$ | $i_{max}$ | $v_i$ |
| --- | --- | --- | --- | --- |
| McLean [3] | 0 | 0 | =1 | 0 ($i > 1$) |
| Fowler-Guggenheim [14] | 0 | $\neq 0$ | =1 | 0 ($i > 1$) |
| Wynblatt-Ku [17, 89] | $\neq 0$ | $\neq 0$ | =1 | 0 ($i > 1$) |
| Wynblatt-Chatain [18] | $\neq 0$ | $\neq 0$ | > 1 | $f_{strain} \cdot \exp[-1.01 \cdot (i-1)^{1.53}]$ |
| Rickman *et al.* [21] | 1/6 | $\neq 0$ | > 1 | $i^{-4}$ |

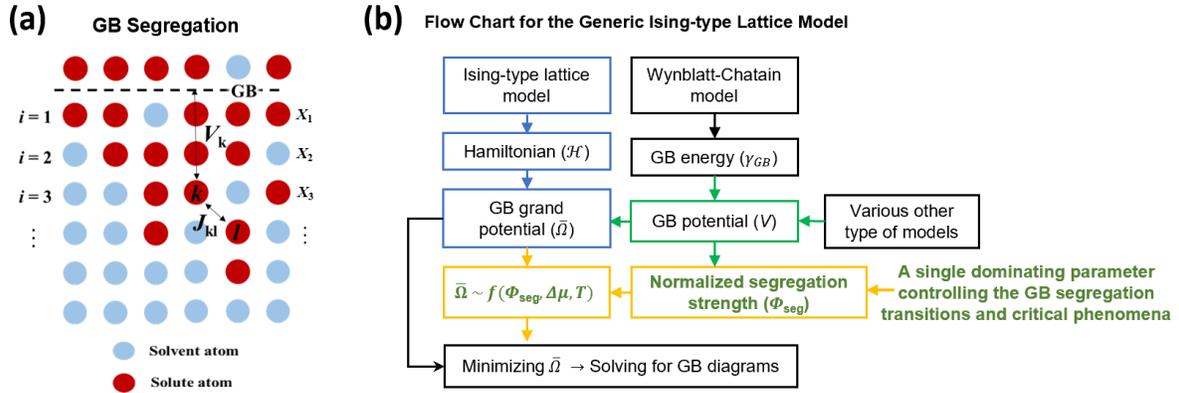

**Figure 1**. **(a)** Schematics of lattice models for grain boundary (GB) segregation. **(b)** The flow chart for developing a generic Ising-type model, which can represent Wynblatt-Chatain and other models via adopting different GB potentials.

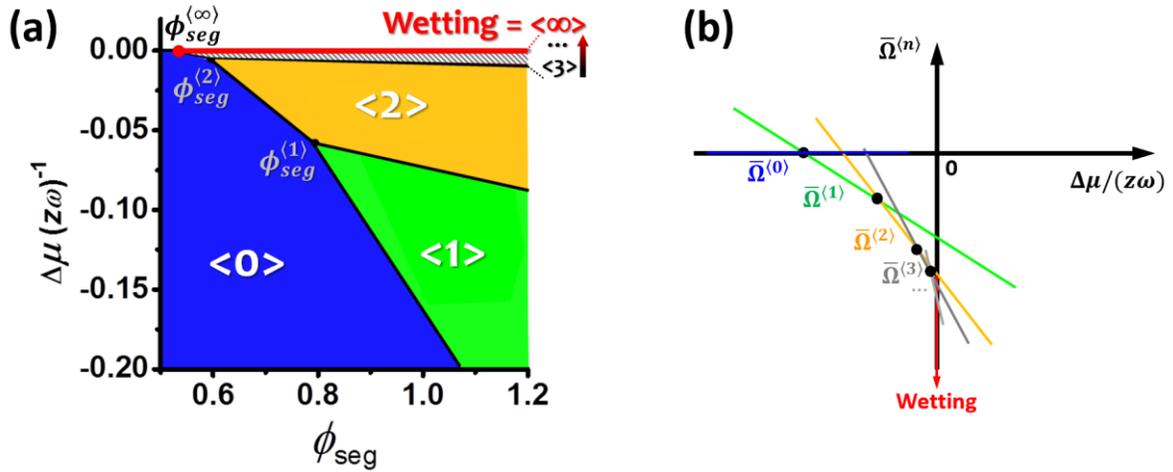

**Figure 2. (a)** Ground GB segregation status at $T = 0$ K as a function of normalized chemical potential difference $\Delta\mu/(z\omega)$ and normalized segregation strength $\phi_{seg}$ (derived from the Wynblatt-Chatain model for a general (100) twist GB, $f_{strain} = 0.5$). The colored regions correspond to different segregation states. Additional 0 K segregation diagrams derived from different models or using different parameters are shown in Suppl. Fig. S2 in the SM. **(b)** Schematic illustration of $\bar{\Omega}^{\langle n \rangle}$ vs. $\Delta\mu/(z\omega)$ curves for a case of strong segregation system at 0 K, showing the formation of a complete series of adsorption states with $n = 0, 1, 2, ...$ to complete wetting.

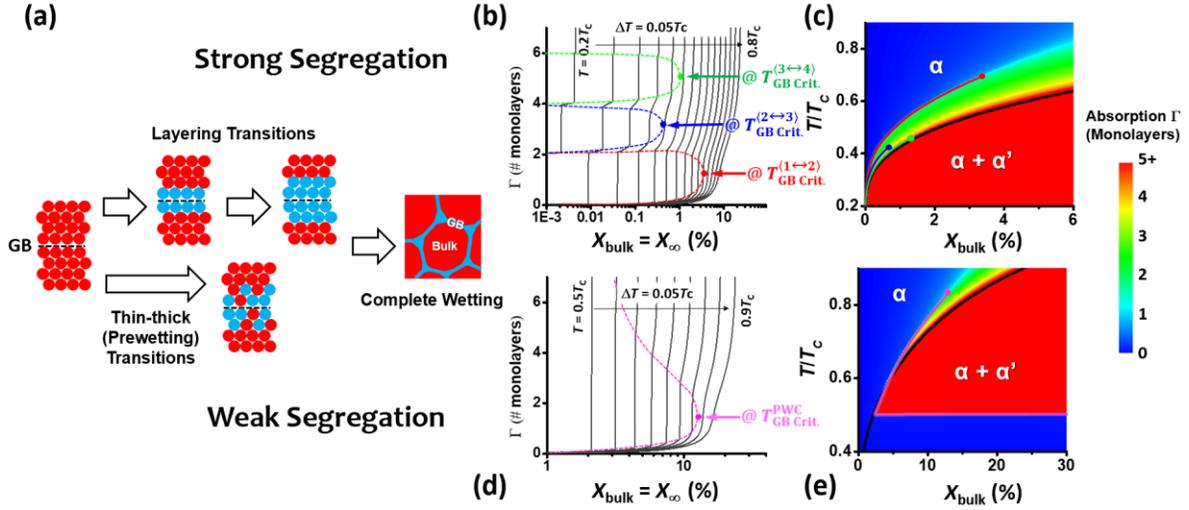

**Figure 3. (a)** Schematic illustration of GB layering *vs.* prewetting transitions. Comparison of calculated GB segregation transitions for **(b, c)** a strong ($\phi_{seg}$ = 1) and **(d, e)** a weak ($\phi_{seg}$ = 0.4) segregation systems. The calculated Γ *vs.* $X_{\text{bulk}}$ curves at various $T/T_C$'s are shown in (b) and (d), where the dash lines denote the states before and after first-order GB transitions and the solid dots represents the GB critical points. The corresponding GB absorption diagrams as functions (maps) of $T/T_C$ and $X_{\text{bulk}}$ are shown in (c) and (e), in which first-order adsorption transition lines are represented by colored lines and bulk phase boundaries are drawn in black lines.

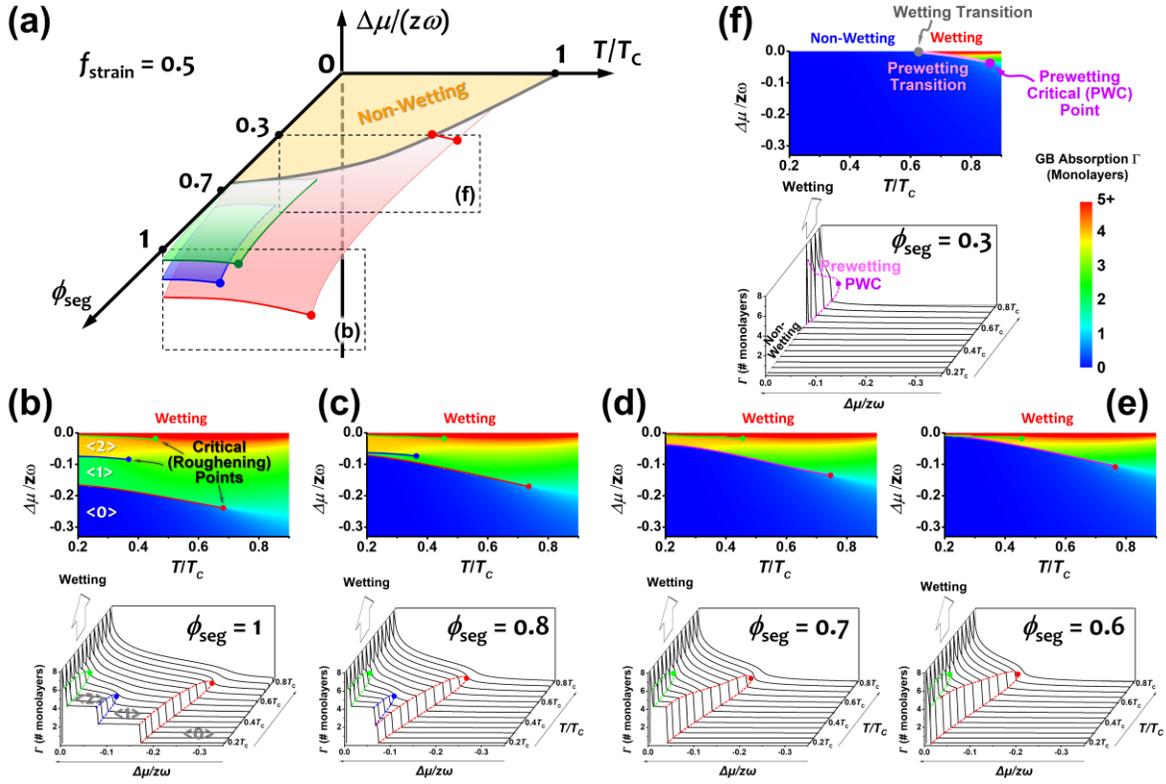

**Figure 4. (a)** A 3D GB segregation transition diagram as a function of $\Delta\mu/(z\omega)$, $T/T_C$, and $\phi_{seg}$. The horizonal $\Delta\mu = 0$ plane represents the bulk two-phase coexistence, below which is the single-phase region of interest. The first-order (bulk) wetting transition line (lying in the $\Delta\mu = 0$ plane) is indicated by the grey solid line, below which is the non-wetting region (colored in yellow). The three hyper surfaces colored in red, blue, and green, respectively, represent the first-order GB segregation transitions. The edges of these adsorption transition hyper surfaces correspond to the critical temperatures (as functions of $\phi_{seg}$). The 2D cross sections of GB segregation diagrams (plotted as functions of $T/T_C$ and $\Delta\mu/(z\omega)$) and the corresponding computed GB adsorption $\Gamma$ vs. $\Delta\mu/(z\omega)$ curves at different $T/T_C$ values for **(b)** $\phi_{seg}$ = 1, **(c)** $\phi_{seg}$ = 0.8, **(d)** $\phi_{seg}$ = 0.7, **(e)** $\phi_{seg}$ = 0.6, and **(f)** $\phi_{seg}$ = 0.3, respectively. The positions of two selected vertical cross sections of (b) for $\phi_{seg}$ = 1 and (e) for $\phi_{seg}$ = 0.3 are denoted in (a) by the dashed rectangles. The vertical 2D cross section at $T = 0$ $K$ of the 3D GB segregation diagram shown in (a) is displayed in Fig. 2(a). More detailed computed results are documented in Suppl. Fig. S5, as well as the slide show of "General GBs Example II", in the SM.

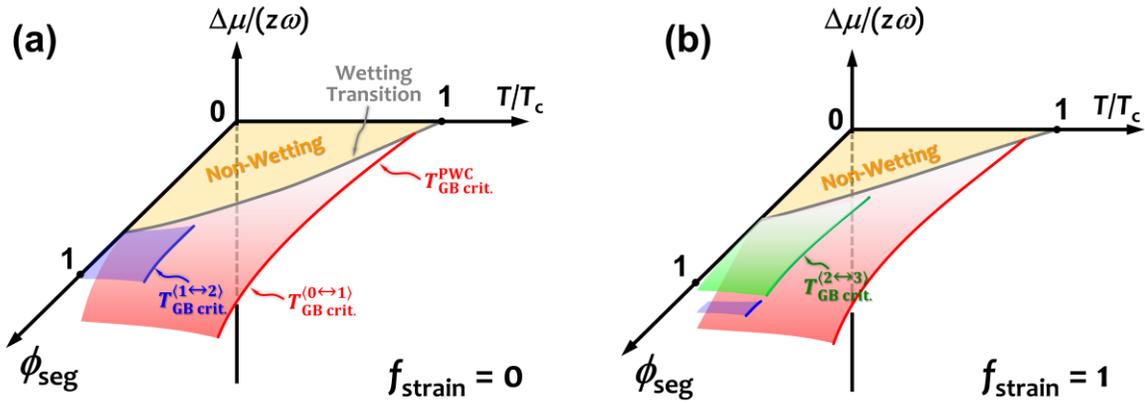

**Figure 5.** Comparison of 3D GB segregation transition diagrams for cases where the segregation is solely driven by **(a)** differential bond energy ($f_{strain} = 0$) or **(b)** strain energy ($f_{strain} = 1$). Note that Fig. 3(a) represents a mixed case (with $f_{strain} = 0.5$) that is in between these two cases. Detailed computed results are shown in Suppl. Fig. S4 and Fig. S6, respectively, as well as the slide show of "General GBs Examples I and III" in the Appendix, in the SM.

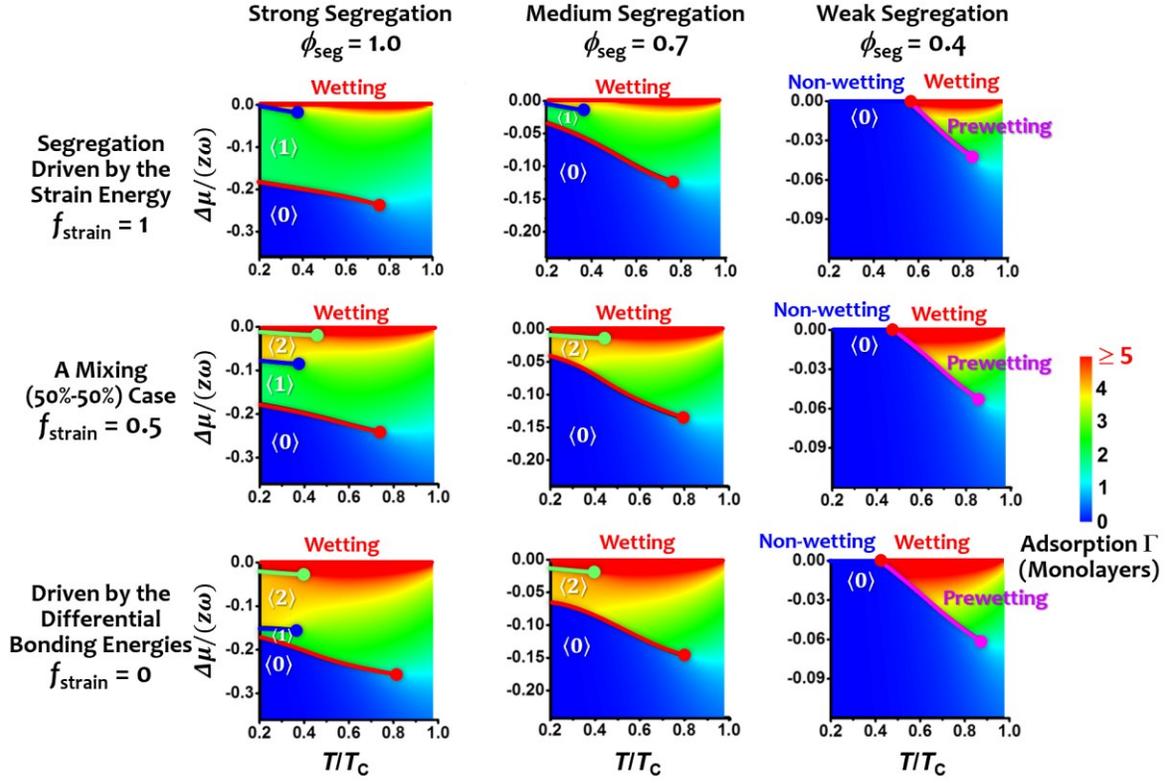

**Figure 6.** Summary and comparison of computed GB adsorption (complexion) diagrams for the three cases of $f_{strain}$ = 0, 0.5, and 1 for representative strong ($\phi_{seg}$ = 1), medium ($\phi_{seg}$ = 0.7), and weak segregation ($\phi_{seg}$ = 0.4) systems (in different scales for $\Delta\mu/(z\omega)$ to clearly show the transitions). A more systematic comparison of 3×9 computed GB segregation diagrams plotted in the same scale can be found in Suppl. Fig. S3 in the SM.

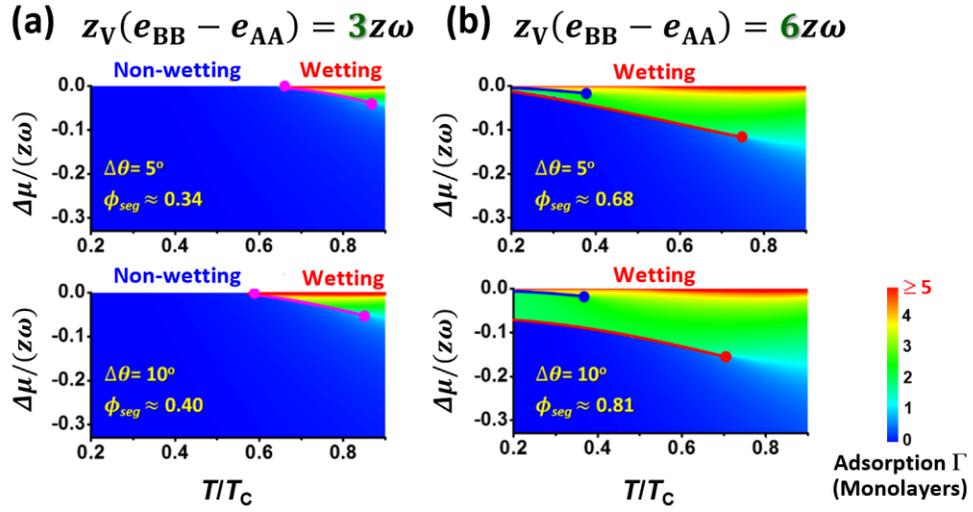

**Figure 7.** Calculated complexion diagrams for small-angle GBs of two different misorientation angles ($\Delta\theta = 5°$ and $10°$, respectively), where the GB segregation is driven by differential bond energies of: **(a)** $z_v(e_{BB} - e_{AA}) = 3z\omega$ and **(b)** $z_v(e_{BB} - e_{AA}) = 6z\omega$.

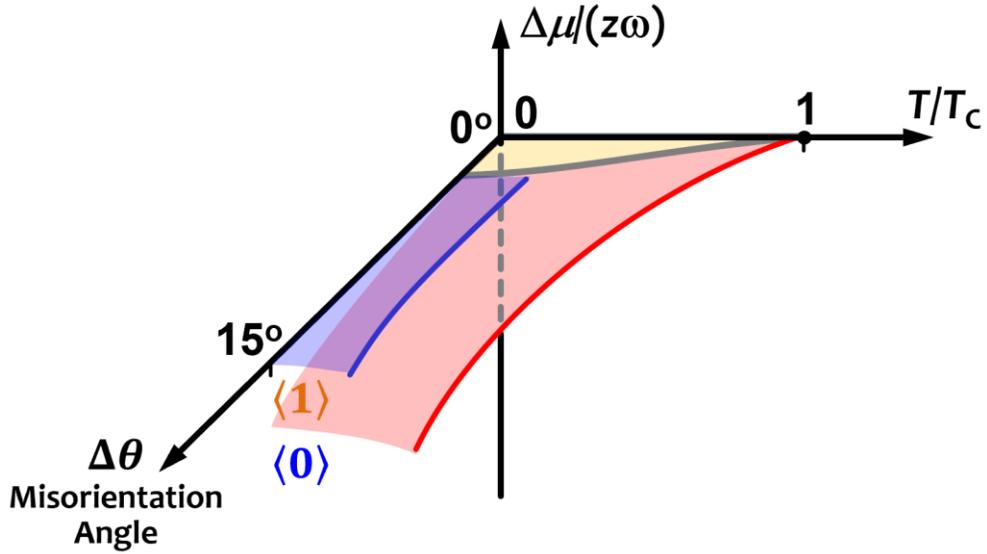

**Figure 8.** Computed 3D GB segregation transition diagram as a function of $\Delta\mu/(z\omega)$, $T/T_C$, and the misorientation angle $\Delta\theta$ for small-angle GBs. Here, the segregation is solely driven by differential bond energies ($f_{strain} = 0$; $z_V(e_{BB} - e_{AA}) = 6z\omega$). See the slide show of "Small-Angle GBs Example I" in the SM for detailed calculated results. Note that this 3D GB segregation diagram is isomorphic to Fig. 5(a) with the following mathematical transformation: $\phi_{seg} \approx 1.08\sin^{1/4}(2\Delta\theta)$ following Eq. (38).

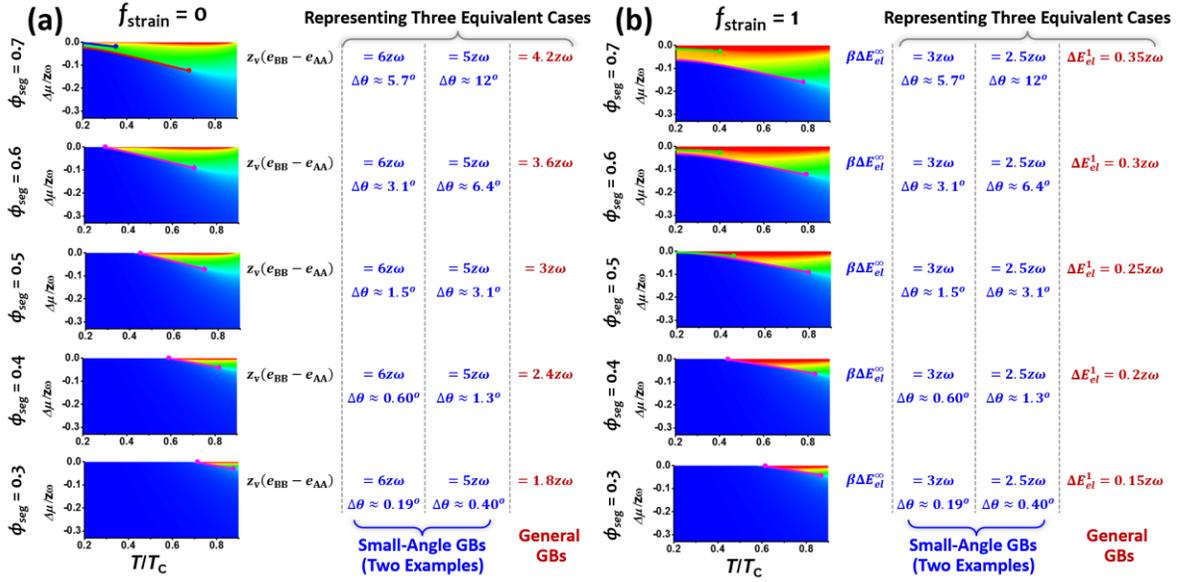

**Figure 9.** A series of computed GB complexion diagrams for **(a)** $f_{\text{strain}} = 0$ and **(b)** $f_{\text{strain}} = 1$ to demonstrate equivalence of small-angle and general GBs with identical normalized segregation strength $\phi_{seg}$. The misorientation angles ($\Delta\theta$) and other model parameters are selected so that the final $\phi_{seg}$ values are identical for two sets of small-angle GBs and a third set of general GBs. Thus, each computed GB segregation diagram can represent three equivalent cases (two small-angle GBs and a general GB).